\documentclass[openacc]{rstransa}

\usepackage{color} 
\newcommand{\corrb}[1]{\textcolor{black}{#1}} 

\titlehead{Research}

\begin{document}

\title{Barchan-barchan and barchan-obstacle interactions: insights from grain-scale studies}

\author{
E. M. Franklin$^{1}$, W. R. Assis$^{2}$, D. S. Borges$^{3}$ and N. C. Lima$^{1}$}

\address{$^{1}$Faculdade de Engenharia Mec\^anica, Universidade Estadual de Campinas (UNICAMP), Rua Mendeleyev, 200, Campinas, SP, Brazil\\
$^{2}$Saint Anthony Falls Laboratory, University of Minnesota, 2 3rd Ave SE, Minneapolis, Minnesota, USA\\
$^{3}$Faculty of Physics, University of Duisburg-Essen, 47057 Duisburg, Germany}

\subject{Earth and Martian surface processes, Landforms, Multiphase flow}


\corres{Erick M. Franklin, ORCID 0000-0003-2754-596X\\
\email{erick.franklin@unicamp.br}}

\begin{abstract}
Sand dunes are bedforms that grow due to the action of a fluid flow over a sand bed or pile. Whenever the fluid flow is mainly in one direction and the availability of sand is limited, crescent-shaped dunes known as \textit{barchans} appear. These dunes are a strong attractor, being found on Earth, Mars, and other celestial bodies, usually in dune fields where they interact with each other, the terrain, and dune-size obstacles. In this review, we discuss the processes and outcomes of the different barchan-barchan and barchan-obstacle interactions, based on \corrb{grain-scale subaqueous} experiments and numerical simulations. We propose that those interactions depend basically on the Shields and Stokes numbers (that are two dimensionless parameters), the transverse position of bedforms, and the size ratio between the interacting objects. In addition, we show in detail the fluid flow, the trajectories of grains, and the resultant force acting on each grain, explaining the mechanisms for the different behaviors observed. Finally, we discuss the implications for the aeolian case, and put into perspective the current findings.
\end{abstract}


\begin{fmtext}
	
\end{fmtext}


\maketitle

\clearpage 
\section{Introduction}

Sand dunes are among the most ubiquitous bedforms observed in nature, covering approximately 6 to 7\% of the Earth's land surfaces \cite{Zheng}, and resulting from the action of a fluid flow over a sand bed or pile. Whenever the flow is roughly unidirectional and the amount of sand is limited (but sufficient), crescent-shaped dunes with tips that point downstream appear \cite{Bagnold_1, Charru_5}. Therefore, these dunes, known as barchan dunes (or simply barchans), emerge as a strong attractor whether in subaqueous or aeolian environments, being found on Earth, Mars, and other celestial bodies \cite{Breed, Hersen_1, Elbelrhiti, Claudin_Andreotti, Bourke, Ewing, Hansen2, Courrech, Cardinale, Courrech2, Love}. Although single isolated barchans can appear, \corrb{crescentic bedforms such as barchans, barchanoids, and other morphologically similar dunes} are usually observed in dune fields where they interact with each other \cite{Hersen_2, Hersen_5, Kocurek, Vermeesch, Genois2, Jiang, Vriend}, the terrain \cite{Finkel, Bourke, Parteli4}, and dune-size obstacles \cite{Breed, Urso, Raffaele, Roback, Jia}.

The morphodynamic evolution of barchan dunes arises from a subtle interplay between the fluid flow, sediment transport, and bedform geometry. When the fluid flows above a given threshold velocity \cite{Bagnold_1, Yalin_2}, it entrains sand as a mobile layer that keeps contact with the fixed part of the sand bed. In the aeolian case, grains move by effectuating ballistic flights with a characteristic length much larger than the grain diameter (known as saltation), while in the subaqueous case grains move by rolling, sliding, and/or small jumps (motion usually called bedload) \cite{Bagnold_1, Raudkivi_1, Yalin_1, vanRijn, Pye}. The moving particles, on the other hand, transfer momentum from the fluid flow to small scale processes such as particle-particle shocks (in the aeolian case) and inter-particle fluid drainage (in the subaqueous case), in the so-called feedback effect \cite{Bagnold_1, Bagnold_2, Bagnold_3, Franklin_9}. The motion of sand engenders erosion in some regions and deposition in others, shaping the granular bed and, again, disturbing the fluid flow \cite{Jackson_Hunt, Hunt_1, Weng, Belcher_Hunt}. In this way, a single barchan dune can be formed from a given granular pile: the fluid-particle-bedform interaction produces an upstream gentle slope, a crescent-shaped downstream face with an inward avalanche slope, and, consequently, two tips (called horns) pointing downstream \cite{Alvarez, Alvarez3}.

The physics at the microscopic scale (grain scale), that is, saltation in aeolian and bedload in subaqueous environments, engenders at the macroscopic scale (bedform scale) barchans that are larger for the former than for the latter environment, but that follow roughly the same dynamics. Therefore, although dunes observed in distinct environments span over different time and length scales, their dynamics is comparable. On Earth, aeolian barchans typically have lengths from tens to hundreds of meters, taking years to displace a considerable distance (their length, for instance), while on Mars they can reach approximately one kilometer in length and take millenniums to move their own length \cite{Hersen_1, Claudin_Andreotti, Parteli2}. In the subaqueous case, the scales in laboratory experiments are typically centimeters and minutes \cite{Franklin_2, Franklin_8, Alvarez}. Despite these differences in size and timescale, the crescent shape remains roughly similar, with sizes in different environments being well described by a relaxation or saturation length $L_{sat}$ \cite{Andreotti_1, Claudin_Andreotti},

\begin{equation}
	L_{sat} = \xi L_{drag} \,\,,
	\label{eq_lsat}
\end{equation}

\noindent where $\xi$ is a proportionality parameter and $L_{drag}$ is the drag length \cite{Hersen_1},

\begin{equation}
	L_{drag} = \frac{\rho_s}{\rho}d  = Sd\,\,
	\label{eq_ldrag}
\end{equation}

\noindent where $\rho_s$ is the grain (sand) density, $\rho$ is the fluid density, $S$ $=$ $\rho_s/\rho$, and $d$ is the grain diameter. The sizes of barchans on aeolian (both terrestrial and Martian) and aquatic environments are, thus, proportional to $L_{sat}$. Given the physical similarities, the onset and evolution of barchans are also commonly described by dimensionless parameters, bearing in mind, however, that details of sand motion are different in aeolian and subaqueous environments. Besides $L_{sat}$, the Shields parameter $\theta$ and the Reynolds number at the grain scale Re$_*$ are of importance. The first represents the ratio between the entraining and resisting forces,

\begin{equation}
	 \theta = \frac{u_*^2}{(S - 1 )gd} \,\,,
	 \label{eq_theta}
\end{equation}

\noindent where $u_*$ is the shear velocity \cite{Schlichting_1} and $g$ $=$ $|\vec{g}|$ is the modulus of the acceleration of gravity. We note that in Eq. \ref{eq_theta} the Shields parameter is written for a turbulent boundary layer. The second parameter indicates the importance of fluid flow inertia at the grain scale,

\begin{equation}
	\mathrm{Re}_* = \frac{\rho u_* d}{\mu} \,\,,
	\label{eq_Re}
\end{equation}

\noindent where $\mu$ is the dynamic viscosity of the fluid. In addition, the shearing caused by the fluid flow needs to surpass a given threshold for entraining the grains. This threshold corresponds to a critical Shields parameter $\theta_{th}$ that is typically plotted as a function of Re$_*$ \cite{vanRijn}. It has been shown, however, that $\theta_{th}$ vary with other parameters and conditions, such as bed armoring \cite{Charru_1}, which can explain the huge dispersion of data when plotted in the $\theta_{th}$--Re$_*$ space. Finally, another parameter that is useful is the Stokes number based on the bulk velocity of the fluid $U$ and the grain diameter $d$,

\begin{equation}
	\mathrm{St} = \frac{U d \rho_s}{18\mu} \,\,,
	\label{eq_St}
\end{equation}

\noindent which quantifies how close the solid particles follow the fluid pathlines \cite{Andreotti_6}.

While a solitary barchan merits consideration due to its rich dynamics and different scales involved, barchans are usually observed in dune fields, where they can interact with each other \cite{Norris, Gay, Elbelrhiti, Vermeesch, Hugenholtz}. From these interactions, barchans exchange mass and redistribute sand within the dune field, regulating their size \cite{Hersen_2, Hersen_5, Kocurek, Genois, Genois2, Assis, Assis2}. Barchan-barchan interactions can also affect the barchan morphology, with the appearance of significant asymmetries \cite{Assis, Assis2}. In addition, Hersen and Douady \cite{Hersen_2} showed that solitary barchans are inherently unstable, tending to grow indefinitely or shrink to disappearance when disturbed from equilibrium, and later Hersen and Douady \cite{Hersen_5} and Robson and Baas \cite{Robson, Robson2} showed that barchan-barchan collisions (when two barchans touch each other) can lead to dune corridors with relatively uniform size distributions, as frequently observed in nature.

Besides dune-dune interactions, field observations show that dunes sometimes approach and interact with topographic features whose dimensions are comparable to the dunes themselves. For example, dunes were observed approaching crater rims on Mars \cite{Breed, Urso, Roback}, houses, buildings, and grasslands and mangroves on Earth \cite{Raffaele, Amaral}, and bridge pillars or similar structures in subaqueous environment \cite{Rubi, Jia}. In the specific case of aeolian dunes on Earth, sand dunes can destroy or swallow human constructions, but the contrary is also true: sometimes human constructions wipe dunes out. Some examples are the degradation of coastal dunes due to houses and other buildings built on the coast. Because many of those dunes protect coastal areas, their elimination engenders the destruction of beaches and threatens cities \cite{Feagin, Martinez}. In those cases, maintaining the existing dune fields is fundamental for preserving vegetated areas, beaches, and biodiversity.

Therefore, understanding how barchans interact with each other or with dune-size obstacles is essential for interpreting dune field patterns, size distributions, spacing, and evolution in both subaqueous and aeolian settings, with implications in engineering, Mars exploration, and nature conservancy. In this review, we compile and critically assess the state of knowledge on barchan-barchan and barchan-obstacle interactions. We aim to provide a multiscale framework that links the various observed behaviors (collision, merging, splitting, bypass, trapping, long-range hydrodynamic coupling) to their controlling physical parameters (e.g., flow strength and unsteadiness; Shields and Stokes numbers; dune-to-dune or dune-to-obstacle size ratios; bedform geometry and relative positions). \corrb{In particular, we focus on recent experiments and numerical simulations at the grain scale}, which unveil the underlying mechanisms in terms of flow perturbations, vortex structures, individual grain mobilization, and resultant forces acting on each grain. Finally, we discuss the implications of these findings for aeolian dune fields, whether on Earth or Mars, pointing out limitations and highlighting key challenges for future research.

\section{Current methods}
\label{section:methods}

The earliest investigations of barchan-barchan interactions were carried out through field observations of aeolian dunes, as, for instance, in Norris and Norris \cite{Norris} and Gay \cite{Gay}. Field studies continue to play a fundamental role in characterizing how barchan dunes interact with each other \cite{Elbelrhiti2, Vermeesch, Hugenholtz}, revealing, for instance, that size regulation within dune fields and the emergence of barchanoid forms are strongly influenced by barchan-barchan collisions. However, because aeolian dunes evolve over long timescales that are typically on the order of decades, time series documenting complete dune-dune interactions are often incomplete. To circumvent these limitations, many of previous studies took advantage of the much smaller and faster scales of subaqueous barchans, carrying out controlled experiments in subaqueous environment or numerical simulations (many of them also in subaqueous environment). Those experimental and numerical studies allowed the investigation of the fundamental mechanisms governing barchan interactions. The same occurs for barchan-obstacle interactions, although in this specific case field measurements are very scarce (to the authors' knowledge, there is no work reporting quantitative field measurements of barchan-obstacle interactions to date). Although in the case of barchan-barchan interactions different numerical methods have been used (for example, agent based methods \cite{Genois, Genois2, Robson, Robson2}), we pay special attention to recent experiments and numerical simulations resolved at the grain scale.

\subsection{Remote sensing}
\label{subsection:remote}

Satellite imagery has been a useful tool for understanding the dynamics of barchan dunes and, more generally, of barchan fields \cite{Norris, Gay, Vermeesch, Elbelrhiti2, Hugenholtz, Azzaoui, Daudon, Jiang}. However, because those images usually concern aeolian environments (both on Earth and on Mars), the timescales are relatively large with respect to the available time series: the very first image taken from Earth by a satellite dates back to 1959 \cite{Harvey}. In particular, in the case of some dune-dune interactions, image sequences of about one century are necessary to observe the complete processes \cite{Vriend} (in some other cases, a few decades are necessary and image sequences capture a large portion of the interaction \cite{Jiang}). Besides the time-series issue, image resolution is sometimes relatively low (mainly in the oldest images), hindering assertions from those measurements only.

Despite those limitations, field measurements were carried out over the last decades \cite{Vermeesch, Elbelrhiti2, Hugenholtz, Azzaoui, Jiang}, showing that the size regulation of barchan dunes, the formation of dune corridors, and the appearance of barchanoid forms are influenced by barchan-barchan interactions. Comparisons of satellite images and field data with numerical and experimental investigations were then conducted over the last years \cite{Zhang, Jiang}, showing that part of the conclusions taken in the laboratory can be applied to the field.

\subsection{Experiments}
\label{subsection:experiments}

To date, the experiments on barchan-barchan and barchan-obstacle interactions were carried out in water channels or tanks. Basically, four types of configurations were used: linear shearing devices, annular shearing devices (Couette-type cells), open-channel flows (water flumes), and pressure driven flows in channels. Endo et al. \cite{Endo2} carried out experiments in a recirculating water flume in which a steady, unidirectional water flow was imposed over a bed containing a pair of sand piles. The piles were initially conical and developed into one barchan dune each, which interacted between themselves while migrating in the flume. Hersen et al. \cite{Hersen_1, Hersen_5} used a tray with grains deposed on its surface, that oscillated linearly and asymmetrically in a water tank in order to create a transient flow with strong shear in only one direction. Under the water flow, the granular piles deformed into barchan dunes that could migrate and interact between them. Bacik et al. \cite{Bacik, Bacik2} used a narrow annular channel filled with water, where the free surface of the water was stirred by paddles, creating a turbulent flow. Under the action of the water flow, a pair of quasi-2D dunes \cite{Bacik} interacted with each other over long times, or a quasi-2D dune interacted with a dune-size obstacle \cite{Bacik2}. Although the experiments of Bacik et al. \cite{Bacik, Bacik2} were not specifically designed for barchan dunes, Assis et al. \cite{Assis7} showed that their quasi-2D dunes  behave similarly to the central slice of a subaqueous barchan.

\begin{figure}[ht]
	\centering
	\includegraphics[width=\linewidth]{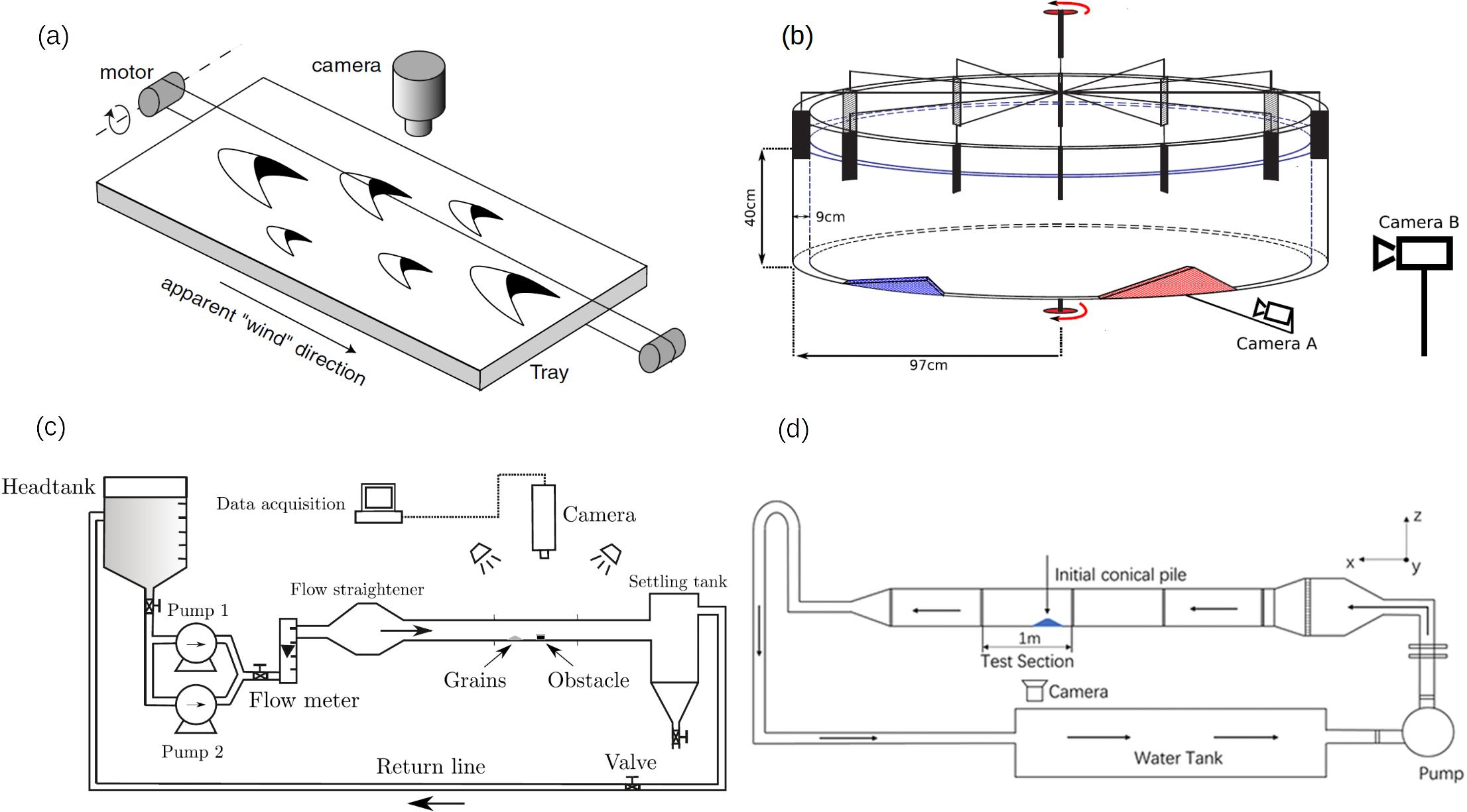}
	\caption{Experimental setups used in investigations of dune-dune or dune-obstacle interactions: (a) Hersen et al. \cite{Hersen_1} (reprinted with permission, Hersen et al. \cite{Hersen_1}, https://doi.org/10.1103/PhysRevLett.89.264301); (b) Bacik et al. \cite{Bacik} (reprinted with permission, Bacik et al. \cite{Bacik}, https://doi.org/10.1103/PhysRevLett.124.054501); (c) Assis et al. \cite{Assis4} (figure extracted from Assis et al. \cite{Assis4}, https://doi.org/10.1029/2023GL104125); (d) He et al. \cite{He} (figure extracted from He et al. \cite{He}, https://doi.org/10.1063/5.0083088).}
	\label{fig:exp_setups}
\end{figure}

Assis et al. \cite{Assis, Assis2, Assis3, Assis4} used a rectangular, closed-conduit water channel of transparent material, in which a pressure driven flow was imposed by means of centrifugal pumps. The channel had an entrance length of more than 40 hydraulic diameters, so that the turbulent flow in the test section was fully developed. In the channel previously filled with water, they placed two initial piles of grains \cite{Assis, Assis2, Assis3} or one initial pile and a dune-size obstacle \cite{Assis4}. By imposing a water flow, the initial piles deformed into barchan dunes that interacted between themselves or with an obstacle, and top-view images acquired with either conventional or high-speed cameras were then post-processed for measuring the desired quantities. He et al. \cite{He, He_2} carried out barchan-barchan experiments similar to those of Assis and Franklin \cite{Assis, Assis2}, but in a water tunnel (the entrance length was less than 11 hydraulic diameters, so that the boundary layer was still evolving in the test section). In the previously described setups, the tank, flume, and channel walls were transparent and measurements were made using cameras and post-processing techniques. Table \ref{tab_exp} shows a summary of the experimental setups and their main dimensions, Tab. \ref{tab_exp2} lists some experimental parameters, and Fig. \ref{fig:exp_setups} shows the layouts ot the four devices used in the mentioned studies.

\begin{table}
	\begin{center}
		\caption{Summary of experimental setups for dune-dune or dune-obstacle interactions and their main dimensions}
		\label{tab_exp}
		\begin{tabular}{l c c}
			\hline\hline
			Work & type & Main dimensions \\
			\hline
			Endo et al. \cite{Endo2} & rectangular flume & 10 m long, 20 cm wide and 50 cm deep\\
			Hersen et al. \cite{Hersen_4, Hersen_5} & moving tray & 150 cm long, 50 cm wide and 20 cm deep\\
			Bacik et al. \cite{Bacik, Bacik2} & annular flume & 88 cm int. and 97 cm ext. radii, 42.4 cm deep\\
			Assis et al. \cite{Assis, Assis2, Assis3, Assis4} & closed-conduit channel & 5 m long, 16 cm wide and 5 cm high\\
			He et al. \cite{He, He_2} & water tunnel & 4 m long, 20 cm wide and 30 cm high\\
			\hline\hline
		\end{tabular}		
	\end{center}
\end{table}

\begin{table}
	\begin{center}
		\caption{Tested conditions for different setups of dune-dune and dune-obstacle interactions: mean diameter $d$, grain density $\rho_s$, order of magnitude of the dune length $O[L]$, Shields number $\theta$, and Reynolds number at the grain scale Re$_*$.}
		\label{tab_exp2}
		\begin{tabular}{l c c c c c}
			\hline\hline
			Work & $d$ & $\rho_s$ & $O[L]$ & $\theta$ & Re$_*$\\
			 & mm & kg/m$^3$ & mm & $\cdots$ & $\cdots$\\
			\hline
			Endo et al. \cite{Endo2} & 0.1 & $\approx$ 2600 & 100 & $\sim$ 0.2$^*$ & $\sim$ 2$^*$\\
			Hersen et al. \cite{Hersen_4, Hersen_5} & 0.15 & 2500 & 100 & N/A & N/A\\
			Bacik et al. \cite{Bacik, Bacik2} & 1.17 & 2500 & 500 & $\sim$ 0.1$^*$ & $\sim$ 30$^*$ \\
			Assis et al. \cite{Assis} & 0.20 and 0.50 & 2500 and 4100 & 10 - 100 & 0.019 - 0.106 & 3 - 10\\
			Assis et al. \cite{Assis2} & 0.20 and 0.50 & 2500 & 10 - 100 & 0.027 - 0.086 & 3 - 7\\
			Assis et al. \cite{Assis3} & 0.20 and 0.50 & 2500 & 10 - 100 &  0.034 - 0.127 & 3-10\\
			Assis et al. \cite{Assis4} & 0.20 and 0.50 & 2500 and 4100 & 10 - 100 & 0.060 - 0.086 & 3 - 8\\
			He et al. \cite{He} & 0.22 & $\sim$ 2500 & 10 - 100 & 0.08 & 2\\
			He et al. \cite{He_2} & 0.25 & 3170 & 10 - 100 & $\sim$ 0.04$^*$ & $\sim$ 2$^*$\\
			\hline\hline
		\end{tabular}
		\\
		\vspace{2mm}
		\small{* Not available in the paper, order of magnitude estimated roughly by the authors.}		
	\end{center}
\end{table}

\subsection{Numerical simulations}

The numerical simulations used to investigate dune-dune interactions are of basically of three types: continuum methods (for the grains), agent-based model, and discrete methods (for the grains). Typically, continuum models \cite{Sauermann_4, Herrmann_Sauermann, Kroy_A, Kroy_C, Kroy_B, Schwammle, Parteli4, Khosronejad} are used for aeolian dunes given the large number of solid particles involved. In many of these methods the fluid shear stresses acting on the dune surface are computed using analytical formulations for the flow perturbed by the topography \cite{Jackson_Hunt, Hunt_1, Weng}, while the granular bed is a considered continuum media whose surface undergoes the fluid stresses. The sediment transport is represented through grains moving predominantly in the longitudinal direction via equations taking into account the momentum transfers between the fluid and grains. Because they are employed mostly for aeolian dunes, those models consider that sand moves predominantly in the longitudinal direction through saltation, with only a small transverse diffusive component due to deviations caused by collisions of salting grains with the bed. Other types of continuum models \cite{Khosronejad} solve large eddy simulations (LES) for the fluid and a suspension layer, given that the most energetic vortices, captured by LES, are responsible for the transport of sand. The coupling with the sediment layer is done by means of mass and momentum conservations. While these continuum simulations have contributed substantially to our understanding of barchan morphodynamics \cite{Sauermann_4, Herrmann_Sauermann, Kroy_A, Kroy_B, Kroy_C, Schwammle, Parteli4, Khosronejad}, they inherently neglect grain-scale behavior and are most appropriate when dunes contain very large numbers of grains.
Other method applied to dune fields is the agent-based approach \cite{Duran5, Worman, Genois, Genois2, Robson, Robson2}. In this framework, each dune is treated as an individual agent characterized by a small set of macroscopic variables, such as size, position, migration velocity, and sand flux, while the surrounding flow and sediment transport are represented implicitly through parameterized flux balances. Dune motion, growth, and decay result from prescribed rules for sand exchange with the bed and from elementary interaction laws describing barchan–barchan encounters, including collision, mass exchange, calving, and merging. Because of their reduced computational cost, agent-based models are well suited for exploring the long-term evolution of dune fields over spatial and temporal scales that are inaccessible to experiments or grain-resolved simulations, having  successfully reproduced large-scale features of dune fields. However, as in continuum models, the underlying dynamics at the microscale is not resolved. Cellular automaton models offer an alternative framework for studying dune–dune interactions by representing the sediment bed on lattices and applying local, rule-based descriptions of erosion, transport, and deposition, without explicitly resolving the grain-scale dynamics \cite{Zhang_3}. This approach has been used for investigating the formation and evolution of dune fields \cite{Zhang_3}, the interaction between 2D dunes \cite{Jarvis}, and barchan-barchan collision \cite{Lin}. Although based on simplified physics, these models reproduce some of the key features of dune–dune interactions.

\begin{figure}[ht]
	\centering
	\includegraphics[width=\linewidth]{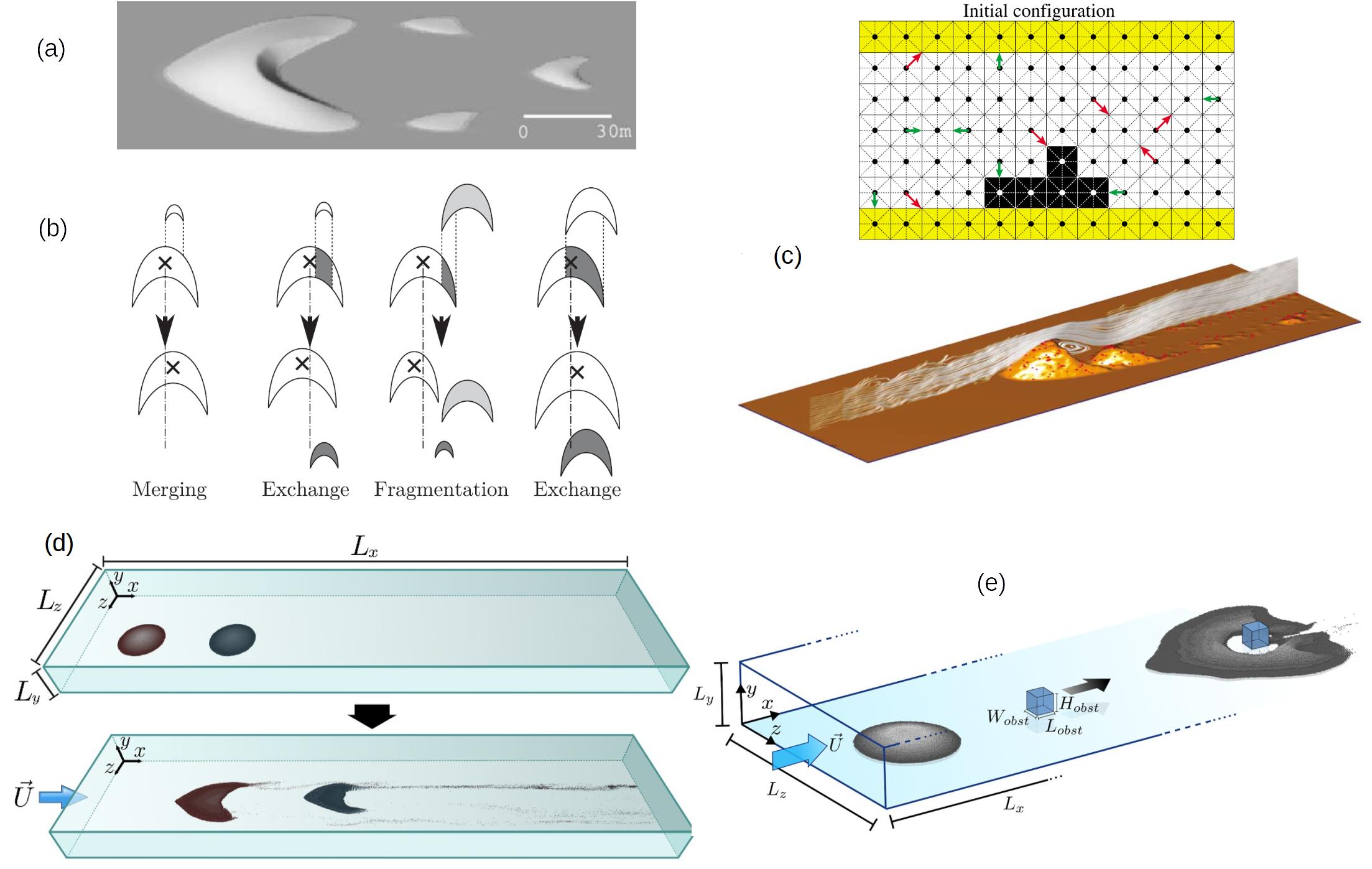}
	\caption{Numerical setups used in investigations of dune-dune or dune-obstacle interactions: (a) Continuous model used in Duran et al. \cite{Duran2} (reprinted with permission, Duran et al. \cite{Duran2}, https://doi.org/10.1103/PhysRevE.72.021308). (b) Agent-based model used in G\'enois et al. \cite{Genois2} (reprinted with permission, G\'enois et al. \cite{Genois2}, https://doi.org/10.1002/grl.50757, 2013). (c) Cellular automaton model of Zhang et al. \cite{Zhang_3} (reprinted with permission, Zhang et al. \cite{Zhang_3}, https://doi.org/10.1029/2009JF001620). (d) CFD-DEM simulations of barchan-barchan interactions of Lima et al. \cite{Lima2} (reprinted with permission, Lima et al. \cite{Lima2}, https://doi.org/10.1029/2024JF007741). (e) CFD-DEM simulations of barchan-obstacle interactions of Lima et al. \cite{Lima3} (figure extracted from Lima et al. \cite{Lima3}, https://doi.org/10.1029/2025JF008504).}
	\label{fig:num_setups}
\end{figure}

In recent years, coupled computational fluid dynamics–discrete element method (CFD–DEM) simulations have been increasingly employed to investigate the morphodynamics of barchan dunes at the grain scale, complementing continuum approaches and laboratory experiments \cite{Alvarez5, Alvarez8, Lima, Lima2, Lima3}. In these models, the fluid phase is computed in an Eulerian framework and the individual grains in a Lagrangian framework, by assuring mass conservation and solving Newton’s laws of motion. In applications involving barchan dunes, the simulations are usually of 4-way type, including particle-particle and fluid–particle interactions (fluid on particles and the feedback of particles on the fluid), and the fluid is water in order to keep the number of particles on the order of 10$^5$. Typically, momentum exchange between the fluid and grains is accounted for through drag, fluid stresses, and added-mass contributions, allowing the evolving granular bed to modify the local flow field while simultaneously computing grain entrainment, transport, and deposition. The main challenges of such simulations are related to their high computational cost, which stems from the modeling strategies required to capture the turbulent boundary layer while employing a spatial discretization fine enough to adequately solve DEM at the grain scale. In this context, current approaches are typically limited to either fully resolved DNS- (direct numerical simulations, more expensive computationally) or unresolved LES-based CFD–DEM formulations. Applications of this methodology to subaqueous barchans have shown that CFD-DEM simulations reproduce the main morphological features of dunes, such as crescentic shapes, horn elongation, and migration rates \cite{Alvarez5, Alvarez8}, as well as mass redistribution during barchan–barchan and barchan–obstacle interactions \cite{Lima, Lima2, Lima3}. In addition, these simulations provide access to quantities that are difficult (or even impossible) to be obtained experimentally, including local flow perturbations, vortical structures, grain trajectories, and the resultant forces acting on individual particles, enabling a detailed analysis of the mechanisms governing dune-dune and dune-obstacle interactions. Although computationally demanding and currently restricted to small dunes and timescales, CFD-DEM approaches explicitly resolve the discrete nature of sediment transport and are, therefore, well suited for studying interaction processes in which grain-scale dynamics play a central role.

\section{Barchan-Barchan interactions}

We begin by stating that we are focusing on binary interactions, that is, when initially only two dunes interact with each other. In this case, basically two types of interaction can occur: (i) when the upstream dune reaches the downstream one and they touch each other (collision); (ii) when the dunes do not collide (but can still exchange mass). Those two behaviors can then be subdivided into subtypes.

\subsection{Subaqueous dunes}

\subsubsection{Monodisperse grains}

Endo et al. \cite{Endo2} and Hersen and Douady \cite{Hersen_5} carried out the first controlled experiments on dune-dune collisions. Endo et al. \cite{Endo2} used a water flume to investigate interactions between aligned barchans, systematically varying the mass ratio of the colliding dunes while keeping the flow rate, initial conditions, and grain properties constant.  They showed that barchan-barchan collisions produce smaller dunes, promoting sand redistribution, and identified three types of collision outcomes: absorption, ejection, and split. These outcomes correspond, in fact, to the merging, exchange, and fragmentation-chasing patterns identified later by Assis and Franklin \cite{Assis}.

Hersen and Douady \cite{Hersen_5} made use of the water tank and tray described in Subsection \ref{section:methods}\ref{subsection:experiments}, which allowed them to investigate collisions between off-centered barchans by varying the transverse separation (often referred to as the impact or offset parameter) while maintaining all other parameters fixed. Their experiments revealed a range of complex collision behaviors that act to redistribute sand between dunes, promoting size-regulation of barchans after repeated interactions. Based on their experimental observations, Hersen and Douady \cite{Hersen_5} proposed a stability framework suggesting that the cumulative effect of dune collisions in a population can counteract the inherent instability of isolated dunes, leading to the emergence of characteristic dune sizes in barchan corridors. In this way, barchan-barchan collisions constitute an active mechanism that shapes field-scale patterns through redistribution of mass among dunes. These insights laid the groundwork for subsequent experimental and numerical studies on dune-dune interactions, and for models that incorporate collision outcomes into descriptions of dune field evolution and size selection (agent-based models).

Bacik et al. \cite{Bacik} showed that quasi-2D dunes in an annular flume can interact over distances much larger than their own length through flow-mediated mechanisms, without requiring direct contact or sediment exchange. Specifically, they observed that a downstream dune can migrate away from an upstream one, giving rise to a long-range repulsive interaction. They attributed this behavior to the turbulent wake generated by the upstream dune, whose coherent flow structures modify the local shear stress and sediment transport experienced by the downstream dune, thereby preventing dunes from touching each other. These findings highlight the role of wake-induced hydrodynamic feedbacks in regulating dune spacing, and provide a new perspective on the self-organization and stability of dune corridors. This was later reinforced by Bacik et al. \cite{Bacik3}, who showed that such wake-mediated equilibrium can be dynamically stable under perturbations (although the point of stability changes according to the problem parameters). We note that the dynamical system proposed by Bacik et al. \cite{Bacik3} should be interpreted with caution when extrapolated to the three-dimensional case. In their study, the system was periodic and the dunes confined by lateral walls, so that their dunes were quasi-2D.

\begin{figure}[ht]
	\centering
	\includegraphics[width=\linewidth]{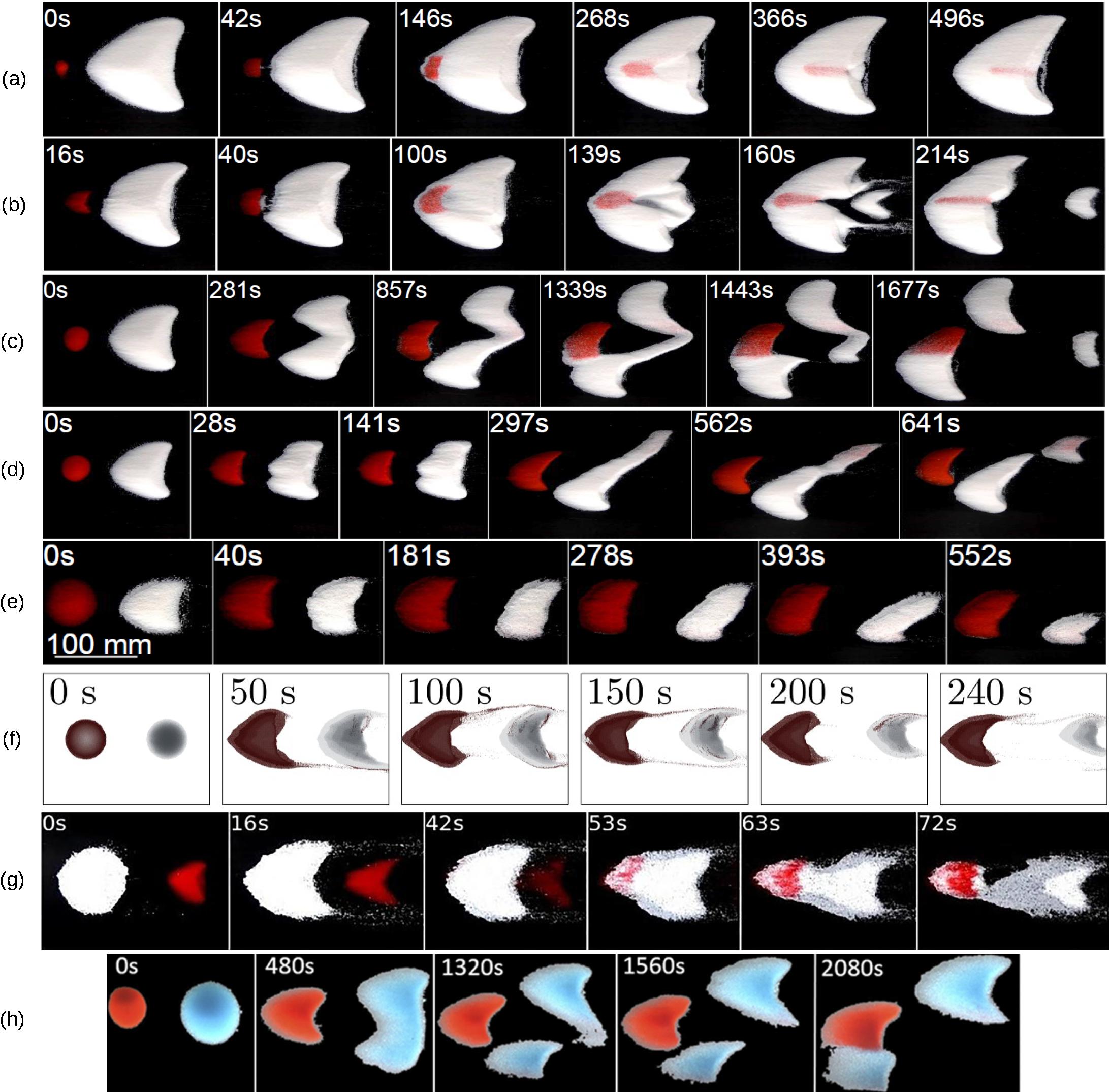}
	\caption{(a-e) Snapshots (top view) of barchan-barchan interactions for aligned dunes, from experiments with monodisperse particles (one peaked distribution). In the snapshots, the water flow is from left to right, the upstream pile consists of red glass beads and the downstream pile of white glass beads, and the corresponding times are shown in each frame. (a) merging; (b) exchange; (c) fragmentation-exchange; (d) fragmentation-chasing; (e) Chasing. (f) Snapshots from numerical simulations of the chasing pattern, showing the grains of each dune at different instants (appearing in maroon for those originally in the upstream barchan and gray in the downstream one). (g) Snapshots of barchan interactions for two initially monodisperse piles of different grains. (h) Snapshots from He et al. \cite{He}, showing a chasing-exchange pattern (note the similarity with the framentation-exchange of panel (c)). In the snapshots, the upstream pile is larger and consists of white spheres with $d$ $=$ 0.5 mm, while the downstream pile consists of red spheres with $d$ $=$ 0.2 mm. Panels (a-e) were extracted from Assis and Franklin \cite{Assis} (reprinted with permission, Assis and Franklin \cite{Assis}, https://doi.org/10.1029/2020GL089464), panel (f) from Lima et al. \cite{Lima2} (reprinted with permission, Lima et al. \cite{Lima2}, https://doi.org/10.1029/2024JF007741), panel (g) from Assis et al. \cite{Assis3} (reprinted with permission, Assis et al. \cite{Assis3}, https://doi.org/10.1029/2021JF006588), and panel (h) from He et al. \cite{He}, https://doi.org/10.1063/5.0083088.}
	\label{fig:dune_dune}
\end{figure}

Assis and Franklin \cite{Assis} conducted an extensive experimental investigation of short-range subaqueous barchan–barchan interactions, in which they generated pairs of barchans with different initial masses, alignments (aligned and off-centered with respect to the flow direction), grain properties (diameter, density, roundness), flow strengths, initial separations, and initial conditions (either conical piles, conical pile and barchan, and barchans). They found that the system evolves independently of the initial conditions, and identified five distinct interaction patterns, listed below in sequence of increasing size of the upstream barchan with respect to the downstream one:

\begin{itemize}
	\item Merging (Fig. \ref{fig:dune_dune}a), when the upstream barchan reaches the downstream one and they simply merge (the size of the upstream barchan is not enough to disturb significantly the fluid flow once collision occurs). 
	\item Exchange (Fig. \ref{fig:dune_dune}b), when the upstream barchan reaches the downstream one and a small barchan is ejected. This happens because the size of the upstream barchan is enough to disturb the fluid flow and, once collision takes place, calve a new barchan that has roughly the size of the upstream one. Although this regime has been compared to solitary waves \cite{Schwammle2}, the barchans merge before ejection takes place and the ejected barchan consists of grains that are different from those of the impacting one, as can be seen in Fig. \ref{fig:dune_dune}b. Therefore, subaqueous experiments produced evidence that, indeed, barchan-barchan interactions cannot be treated as solitary waves \cite{Livingstone}.
	\item Fragmentation-exchange (Fig. \ref{fig:dune_dune}c), when the downstream dune begins splitting into two smaller dunes (due to flow disturbances caused by the upstream barchan). The upstream barchan then reaches the splitting dune, and a small barchan is ejected once they touch each other.
	\item Fragmentation-chasing (Fig. \ref{fig:dune_dune}d), when the flow disturbances created by the upstream barchan are such that the downstream dune splits into two smaller dunes before being reached by the upstream barchan. The new smaller dunes outrun the upstream barchan, and they never touch each other.
	\item Chasing (Figs. \ref{fig:dune_dune}e and Fig. \ref{fig:dune_dune}f), when the upstream barchan neither reaches nor splits the downstream one. This happens when the two barchans initially have roughly the same size, and it is common that the flow disturbances due to the upstream barchan erode considerably the downstream one. When that happens, the downstream barchan outruns the upstream one and their initial separation increases.
\end{itemize}

\begin{figure}[ht]
	\centering
	\includegraphics[width=\linewidth]{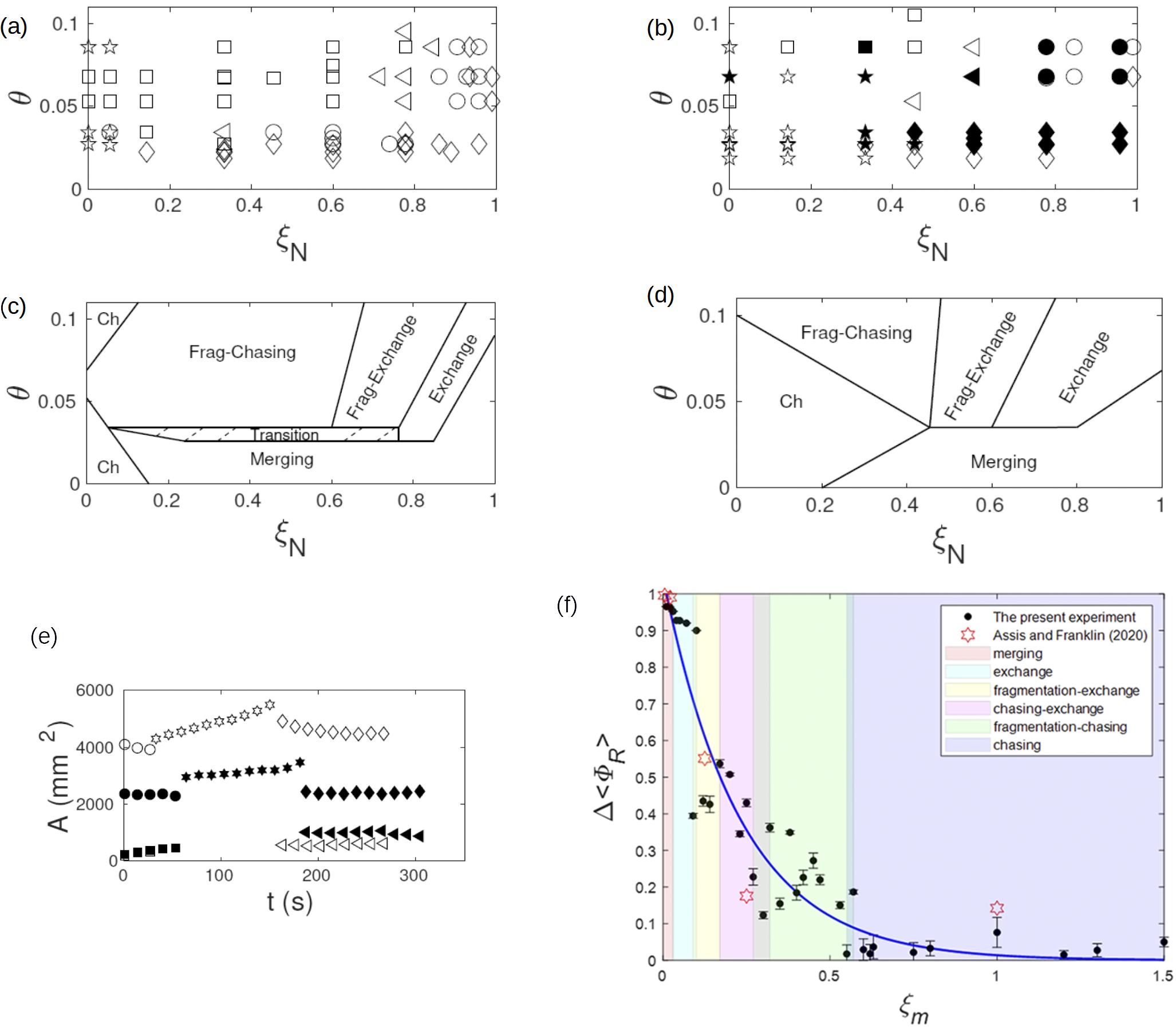}
	\caption{(a-b) Patterns of barchan-barchan interactions as functions of $\xi_N$ and $\theta$ for (a) aligned and (b) off-centered barchans. Stars, diamonds, circles, squares and triangles correspond to chasing, merging, exchange, fragmentation-chasing and fragmentation-exchange, respectively. In panel (b), open symbols correspond to $\sigma$ $<$ 0.5 and solid symbols to $\sigma$ $\geq$ 0.5. (c-d) Boundaries between different patterns in the $\theta$--$\xi_N$ space for the aligned and off-centered barchans, respectively, where Ch stands for chasing and Frag to fragmentation. (e) Area variation along time for the exchange pattern. Squares and circles correspond to the initial upstream (impact) and downstream (target) barchans, respectively, stars to the merged bedform, and diamonds and triangles to the merged bedform and new (expelled) barchan, respectively. Open symbols correspond to the aligned and solid symbols to off-centered cases. (f) Representative degree of occupation $\Delta \left< \Phi_R \right>$ (subtraction of the initial value from the averaged value $\left< \Phi_R \right>$ obtained for the final state) as a function of the mass ratio $\xi_m$ between the upstream and downstream barchans. The graphic shows a monotonic variation and good organization of data, but no other parameter than $\xi_m$ was varied in the data shown. Panels (a-e) were extracted from Assis and Franklin \cite{Assis} (reprinted with permission, Assis and Franklin \cite{Assis}, https://doi.org/10.1029/2020GL089464), and panel (f) from He et al. \cite{He}, https://doi.org/10.1063/5.0083088.}
	\label{fig:dune_dune2}
\end{figure}

\noindent These patterns emerge in both aligned and off-centered configurations, showing that they result from a combination of dune size (with which they vary monotonically), fluid forcing, and grain transport. Therefore, the authors proposed \textit{ad hoc} classification maps that depend on the alignment of barchans, Shields number ($\theta$), and ratio between the number of grains of each dune ($\xi_N$),

\begin{equation}
	\xi_N = \frac{\Delta_N}{\Sigma_N}
\end{equation}

\noindent where $\Delta_N$ is the difference in the number of grains forming each initial pile and $\Sigma_N$ is the sum of those numbers. The maps, shown in Figs. \ref{fig:dune_dune2}a and \ref{fig:dune_dune2}c for aligned and in Figs. \ref{fig:dune_dune2}b and \ref{fig:dune_dune2}d for off-centered barchans, organize reasonably well the observed patterns. In the off-centered case, open symbols correspond to $\sigma$ $<$ 0.5 and solid symbols to $\sigma$ $\geq$ 0.5, where $\sigma$ \cite{Hersen_5} is the off-center parameter (two times the transverse separation divided by the sum of widths of upstream and downstream barchans \cite{Assis}). In particular, Assis and Franklin \cite{Assis} also showed that in the exchange case the size of the ejected barchan is roughly the same of that of the impacting one, as shown in Fig \ref{fig:dune_dune2}e. Later, Assis and Franklin \cite{Assis2} tracked the motion of individual grains as dunes approached and exchanged mass (in both colliding and non-colliding cases), obtaining the trajectories of grains moving from one barchan to another. In merging and exchange interactions, they showed that grains from the upstream (impacting) barchan exhibit a diffusion-like spreading component when moving over the downstream (target) dune, so that mass exchange can be modeled by stochastic processes. By quantifying the motion of individual grains, the study provided statistical data of displacement distances and velocities, revealing how mass transfer and grain redistribution occur at the smallest scales during interactions that ultimately shape a dune field.

He et al. \cite{He} revisited the experiments of barchan-barchan interactions of Assis and Franklin \cite{Assis, Assis2} in their water tunnel, but keeping all parameters constant except the mass ratio between the dunes, and carried out an analysis using data from both experimental setups. They obtained patterns that are similar to those proposed by Assis and Franklin \cite{Assis} plus a transitional regime that we add to the previous list:

\begin{itemize}
	\item Chasing-exchange (Fig. \ref{fig:dune_dune}h), which is a transitional regime between the fragmentation-chasing and fragmentation-exchange patterns.
\end{itemize}	
	
\noindent Table \ref{tab_nomenclature} summarizes the nomenclatures adopted in different studies and their corresponding equivalences. 

\begin{figure}[ht]
	\centering
	\includegraphics[width=\linewidth]{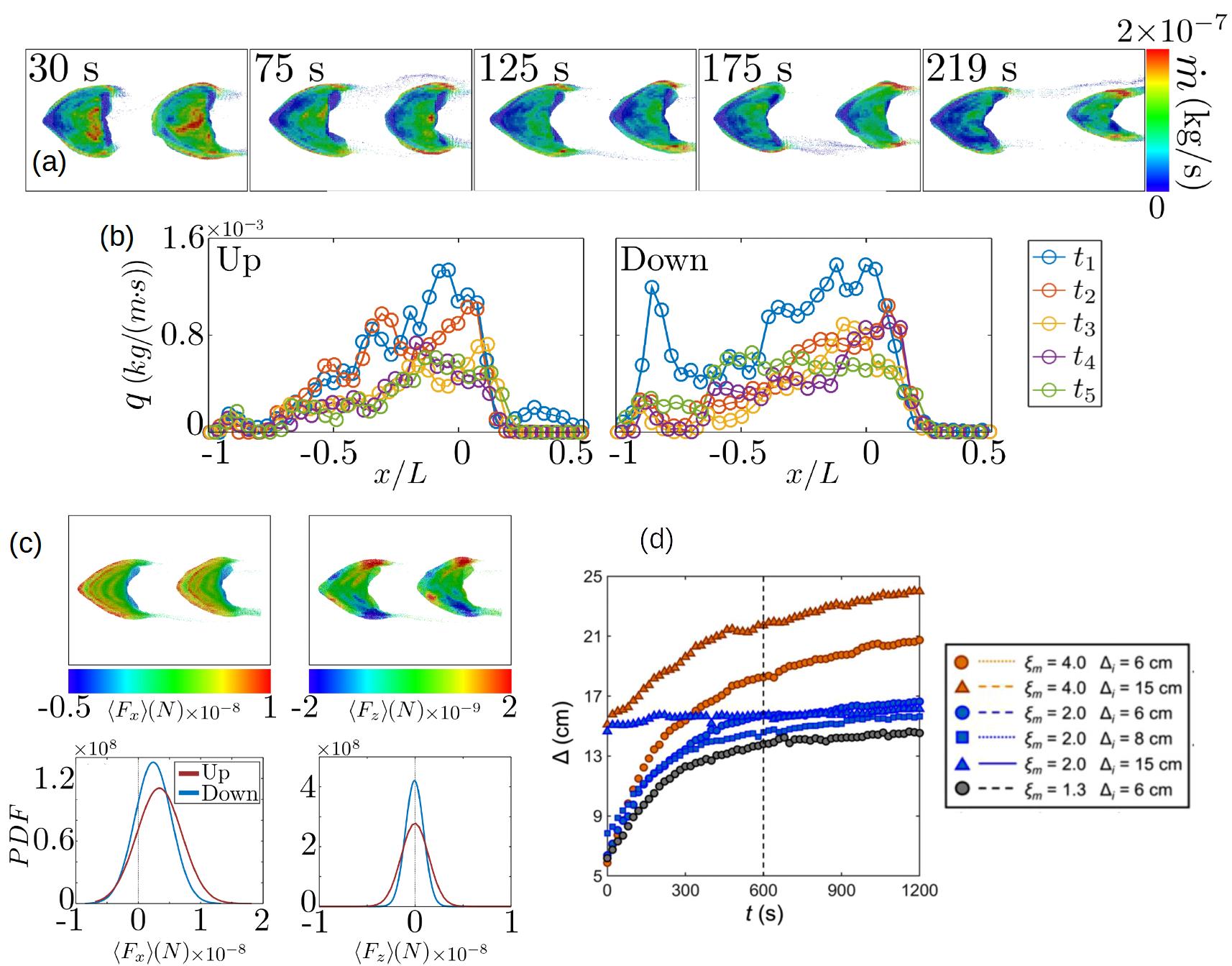}
	\caption{(a) Snapshots showing the grains of each dune (top view) at different instants, colored in accordance with the average mass flow rate $\dot{m}$ of the region they are in. The averages are computed in intervals $t_1$ to $t_5$ ($t_1$ = 10--50 s, $t_2$ = 51--100 s, $t_3$ = 101--150 s, $t_4$ = 151--200 s, $t_5$ = 201--238 s). (b) Sediment flux $q$ along the barchan (computed based on its central slice only, $\approx$ 1.2 mm thick), averaged over the $t_1$ to $t_5$ intervals, for the upstream (Up) and downstream (Down) barchans. (c) Resultant force on the grains of dunes in aligned configuration. From left to right, top to bottom: top view of dunes colored in accordance with the average resultant force in each region, on the longitudinal $\left< F_x \right>$ and transverse $\left< F_z \right>$ directions; histograms for the of longitudinal $\left< F_x \right>$ and transverse $\left< F_z \right>$ components of forces. (d) Longitudinal separation between barchans as a function of time, parameterized by the mass ratio $\xi_m$ and the initial separation $\Delta_i$. Panels (a-c) were extracted from Lima et al. \cite{Lima2} (reprinted with permission, Lima et al. \cite{Lima2}, https://doi.org/10.1029/2024JF007741), and panel (d) from He et al. \cite{He_2} (reproduced from He et al. \cite{He_2}, Phys. Fluids, 35, 106609, 2023, https://doi.org/10.1063/5.0169485, with the permission of AIP Publishing).}
	\label{fig:dune_dune3}
\end{figure}

In addition, He et al. \cite{He} introduced two scalar measures: a degree of diffusivity $\Phi_B$ and a degree of occupation $\Phi_R$. Those dimensionless parameters, valid in principle only when all parameters other than dune sizes remain constant, collapse barchan-barchan outcomes onto smooth trends, as shown in Fig. \ref{fig:dune_dune2}f. With that, they proposed that repeated collisions drive the system toward chasing-dominated configurations. Later, He et al. \cite{He_2} focused on the chasing pattern and, among other results, showed that the inter-dune space tends to a constant value in the chasing case (as shown in Fig. \ref{fig:dune_dune3}d, for example). They also proposed a model for explaining the convergence of inter-dune spacing, based on the balance between wake-induced repulsion, an embracing vortex acting on the downstream dune, and size-dependent migration speeds.

\begin{table}
	\begin{center}
		\caption{Different nomenclatures adopted in the literature for the outcomes of barchan-barchan interactions. Each row correspond to the same pattern.}
		\label{tab_nomenclature}
		{\color{black}
			\begin{tabular}{l l l l}
				\hline\hline
				Endo et al. \cite{Endo2} & Duran et al. \cite{Duran5} & Assis and Franklin \cite{Assis, Assis2} & He et al. \cite{He}\\
				\hline
				Absorption & Coalescence & Merging & Merging\\
				Ejection & Solitary wave & Exchange & Exchange\\
				-- & \corrb{Breeding, Budding} & Fragmentation-exchange & Fragmentation-exchange\\
				Split & \corrb{--} & -- & Chasing-exchange\\
				Split & -- & Fragmentation-chasing & Fragmentation-chasing\\
				-- & -- & Chasing & Chasing\\
				\hline\hline
			\end{tabular}
		}
	\end{center}
\end{table}

More recently, Lima et al. \cite{Lima2} addressed the chasing pattern (barchan–barchan repulsion) using CFD-DEM (grain-scale simulations). As in the experiments of Assis and Franklin \cite{Assis, Assis2}, a pair of granular heaps was allowed to evolve into interacting barchan dunes under an imposed water flow, with both the fluid motion and individual grain dynamics solved at relatively high spatial and temporal resolutions. The simulations provided detailed measurements of particle positions, velocities, resultant forces, and mass balances for each dune throughout their interaction. Lima et al. \cite{Lima2} found that the downstream barchan tends to shrink faster than the upstream one because of a disturbed flow that enhances erosion on the downstream dune, mainly in the first stages of dune-dune interaction (as shown in Figs. \ref{fig:dune_dune3}a and \ref{fig:dune_dune3}b). This erosion and mass redistribution explain why, in the chasing pattern, the downstream barchan can maintain a relatively high migration speed or even outrun the upstream one without dune-dune collision, offering a mechanistic explanation for that pattern. Finally, Lima et al. \cite{Lima2} showed that the magnitudes of the resultant force acting on each grain of both dunes are similar (as shown in Fig. \ref{fig:dune_dune3}c), so that the higher erosion of the downstream barchan is due to a higher number of moving grains.

\subsubsection{Bidisperse grains}

Most of previous studies on dune-dune interactions focused on monodisperse grains, yet natural dunes are composed of polydisperse sediments with a given range of sizes and densities. In this context, Assis et al.\cite{Assis3} carried out experiments in their water channel (described in Subsection \ref{section:methods}\ref{subsection:experiments}), in which pairs of dunes were formed from either bidisperse mixtures of grains or two distinct monodisperse grain types. Thus, each test corresponded to a pair of dunes, with initially either bidisperse grains or different monodisperse grains, that interacted with each other. The morphologies, dynamics, and grain distribution of dunes were measured by image processing along the dune-dune interactions. The authors chose bidisperse distributions for two reasons: (i) start understanding key features of polidispersity on barchan-barchan interactions, and (ii) understand the problem when two main particle diameters are present in the field. For example, dunes with two distinct colors were observed on the north pole of Mars and have been associated with the presence of coarse-grained ice together with sandy sediments \cite{Schatz, Feldman, Pommerol}.

The authors found that grain size distribution and relative concentration influence the patterns and outcomes of dune-dune interactions. For bidisperse barchans (with two grain sizes mixed within each dune), the interaction sequence often produced a transient stripe on the dune surfaces that reflects grain sorting during motion, a behavior analogous to the transverse stripes observed previously in single bidisperse barchans \cite{Alvarez6}. When each dune consisted of a distinct (between themselves) monodisperse grain type, the interaction patterns diverged significantly from the classical monodisperse case: depending on relative sizes and grain properties, dune collisions could result in scenarios where the upstream dune is larger than the downstream one at impact, a configuration that contradicts many prior assumptions based on monodisperse experiments. This is, indeed, the case shown in Fig. \ref{fig:dune_dune}g, reproduced from Assis et al. \cite{Assis3}: the authors carried out one test in which a larger upstream barchan reached and collided with a smaller one, the key being that the larger dune consisted of larger grains, so that it migrated faster than the smaller dune.

In addition to characterizing new interaction regimes, Assis et al. \cite{Assis3} proposed a timescale for the cases of barchan-barchan collisions (merging, exchange, and fragmentation-exchange types) that applies not only to the bidisperse and two-species monodisperse cases they investigated, but also to traditional monodisperse interactions in subaqueous environment. By considering the timescale $t_s$ as the typical time for barchan collisions, the time is given by dividing the initial separation $\Delta x_d$ of barchans by their relative velocity $\Delta V_d$ (difference between celerities of the impact and target barchans),

\begin{equation}
	t_{s} = \frac{\Delta x_d}{\Delta V_d} \,\,.
	\label{Eq:timescale}
\end{equation}

\noindent In its turn, the celerities of barchans of different grain types and sizes, under different water velocities, was found by Franklin and Charru \cite{Franklin_8} to depend on all those parameters (at least in the ranges 1 $\leq$ $Re_*$ $\leq$ 11 and 0.02 $\leq$ $\theta$ $\leq$ 0.24). By inserting the expression proposed by Franklin and Charru \cite{Franklin_8} in Eq. \ref{Eq:timescale}, Assis et al. \cite{Assis3} found that

\begin{equation}
	t_{s} = \frac{\Delta x_d S}{u_*} \left| \frac{\bar{d_t}}{D_t} - \frac{\bar{d}_i}{D_i}\right|^{-1}
	\label{Eq:timescale2}
\end{equation}

\noindent where $\bar{d}$ $=$ $\left( \phi_{1}/d_1 + \phi_{2}/d_2 \right) ^{-1}$ and $\bar{\rho}_s$ = $\phi_1 \rho_1 + \phi_2 \rho_2$ are the mean diameter and density \cite{Alvarez6}, respectively, of grains of species 1 and 2, $\phi_j$ is the percentage (mass basis) of the species $j$ in the mixture, and the subscripts $i$ and $t$ correspond to the impact and target barchans. In the case of monodisperse dunes, $\bar{d}$ $=$ $d$ and $\bar{\rho}_s$ $=$ $\rho_s$. 

When the measured duration is divided by $t_s$, Assis et al.\cite{Assis3} found that the normalized value varies between 0.04 and 2. The timescale $t_s$ provides a scaling framework by which the duration of interactions can be compared across different sediment compositions, and, although valid for subaqueous dunes, it offers a potential bridge toward interpreting interaction durations in aeolian fields.

\subsection{aeolian dunes}

We discuss now a few studies on barchan-barchan interactions in aeolian environment that used remote sensing (mainly aerial and satellite images). As stated in Subsection \ref{section:methods}\ref{subsection:remote}, time series are limited given the large timescales of the interactions involved. Still, some authors could identify image sequences that show a considerable portion of the interplay between dunes along time. In addition, the first observations on dune-dune interactions were based on field measurements of aeolian barchans \cite{Norris, Gay}, and, therefore, we review briefly a few of them next.

Early field studies showed that the migration rates of dunes scale inversely with their sizes \cite{Bagnold_1}, so that the differences in migration velocities lead to frequent dune-dune interactions. Therefore, collisions are not rare anomalies, but persistent processes that regulate dune-field organization: smaller and faster dunes overtake larger ones and trigger coalescence, exchange, or fragmentation, which results in corridors of size-selected barchans \cite{Hersen_5}. Numerical models and experiments with subaqueous barchans reproduce, thus, the main behaviors observed in the field \cite{Zhang}, including complex exchanges and calving processes \cite{Vermeesch, Hugenholtz, Worman}.

\begin{figure}[ht]
	\centering
	\includegraphics[width=0.85\linewidth]{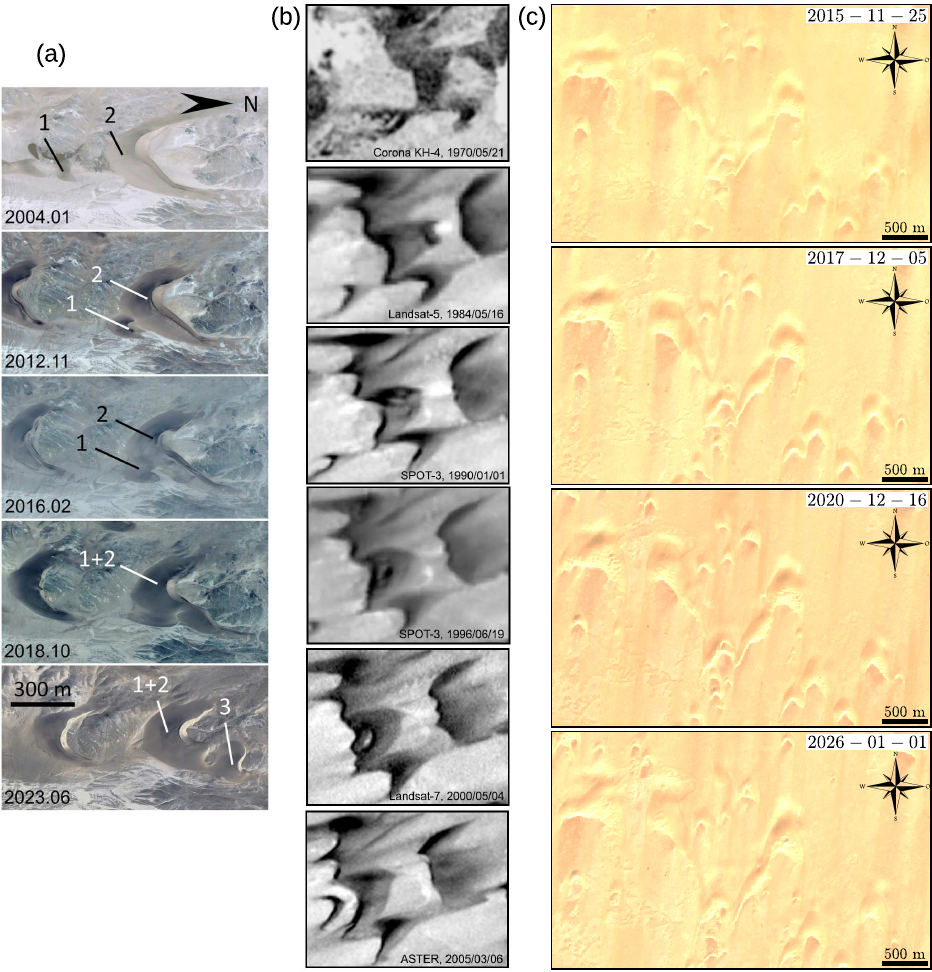}
	\caption{(a) Snapshots with remote-sensing photographs (26$^\circ$52' S, 15$^\circ$20' E) showing two dunes under an off-centered exchange process. In the images, the wind blows from left to right, and the dates (year.month) are indicated in each panel. Figure extracted from Zhang et al. \cite{Zhang}, https://doi.org/10.1029/2023JF007617, reprinted with permission. (b) Satellite images (16$^\circ$48'27'' N, 17$^\circ$18'41'' E, Bod\'el\'e Depression in Chad) showing two barchans undergoing an aligned exchange. In the images, the wind blows from the top-right to the bottom left corners, and the dates and satellites used are indicated in each panel. Figure extracted from Vermeesch \cite{Vermeesch}, https://doi.org/10.1029/2011GL049610, reprinted with permission. (c) Snapshots of satellite images showing part of a barchan field in the Sahara desert, in which barchans undergo complex interactions. The field is located on 21.44$^\circ$ latitude, -16.76$^\circ$ longitude (center of image), the images are from the Sentinel-2-L2A satellite, and the dates are shown in each panel. Courtesy Copernicus EU \cite{Supplemental_Copernicus}.}
	\label{fig:dune_dune4}
\end{figure}

Vermeesch \cite{Vermeesch} analyzed image sequences corresponding to a period of 45 years of an exchange process taking place in the Bod\'el\'e Depression in Chad. The author observes that  a mass transfer takes place during the collision between two barchans, although from the images (Fig. \ref{fig:dune_dune4}b, for example) they seem to interact as solitary waves (that is, to travel one through the other). Hugenholtz and Barchyn \cite{Hugenholtz} used image sequences of approximately 10 years to study two sites along the coastal plain of Namibia. They showed examples of the exchange pattern, and proposed that the ejection of the new barchan is due to calving (erosion of part of the target barchan by flow disturbances caused by the impacting one), as shown later by Worman et al. \cite{Worman} in the aeolian case and Assis and Franklin \cite{Assis, Assis2} in the subaqueous case (and also confirming the mass exchange observed by Vermeesch \cite{Vermeesch}). More recently, Jiang et al. \cite{Jiang} made use of high-resolution satellite images to study merging and exchange/fragmentation patterns in the southeastern edge of the Taklimakan Desert in China. They identified lateral, offset, and coaxial types of merging, as well as several distinct fragmentation pathways including lateral fragmentation and horn detachment. They also found that the relationship between wind forcing and dune height is not the same for dunes that are isolated, merging, and fragmenting, indicating that wind dynamics modulate interaction outcomes. Finally, they showed that the alignment of barchans influences the thresholds for merging and fragmentation under unidirectional winds. These results are in agreement with most of the the results obtained for subaqueous barchans \cite{Assis, Assis2, He, Lima2}, and are generally included in recent agent-based models \cite{Robson, Robson2}. Finally, Fig. \ref{fig:dune_dune4}c shows image sequences of barchans in a dune field on the Sahara desert undergoing barchan-barchan interactions. The sequences illustrate that it is common to have complex interactions, in which three or more barchans interact between themselves.

Agent-based models are very important and useful for computing the evolution of large barchan fields over long timescales, since they treat dunes as discrete entities interacting through simplified rules for migration, sand exchange, collisions, and calving \cite{Duran5, Worman, Robson, Robson2}. Therefore, these methods are ideal for simulating the behavior of fields of aeolian dunes. Using an agent-based model including sand-flux exchange and collisions, Durán et al. \cite{Duran5} showed that collisions can contribute to size selection in barchan fields, reproducing features such as corridors of dunes that increased in size with the downwind distance. Later, Worman et al. \cite{Worman} proposed that dune calving plays a major role in regulating dune sizes and in the emergence of field-scale structures, showing that interactions through horn leakage and calving can lead to patches and corridors resembling those observed in nature. More recently, Robson and Baas \cite{Robson} developed a two-flank agent-based model capable of reproducing all known barchan interaction patterns, dune asymmetry, and calving, while Robson and Baas \cite{Robson2} employed the same model to simulate large swarms of barchans. They showed that the interplay between collisions, calving, and sand-flux exchange can produce longitudinally homogeneous dune fields and realistic swarm structures, similar to the corridors of size-selected dunes observed in nature. Their simulations also showed that changes in environmental conditions, such as bimodal winds and variations in sand supply, propagate through the swarm and affect dune asymmetry, density, and size distributions.

Cellular automaton methods have also been employed in aeolian dune-dune interactions. For example, using the three-dimensional cellular automaton model ReSCAL, Lin et al. \cite{Lin} investigated the effects of the dune-size ratio and interdune spacing on binary barchan interactions. They showed that the outcome of collisions depends not only on the size ratio between the dunes, but also on the interdune spacing, proposing a transition surface separating coalescence and ejection regimes as functions of those parameters. Their simulations reproduced qualitatively field observations and subaqueous experiments. In addition, because their model includes dune calving, Lin et al. \cite{Lin} showed the role of sand loss and secondary dune production in the evolution and size regulation of barchan fields, shedding light on dune-dune interaction in aeolian environment.

Recently, Zhang et al. \cite{Zhang} reviewed the similarities between aeolian barchans and their downsized subaqueous counterparts, showing evidences that similarities in morphology and dune-dune interactions persist despite major differences in the particle-scale transport mechanisms. In particular, they showed that in the formation of single barchans the resulting morphologies share a high degree of similarity in both environments, while for barchan-barchan interactions the degree of similarity decreases as the interaction complexity increases: in a process involving collision with subsequent exchange, similarities decrease as the process evolves from initial merging, ejection of the new barchan, and subsequent outrun of the ejected barchan. The authors conclude that the initial volume (or mass) ratio between dunes is not the only parameter controlling barchan-barchan interactions, and that morphological, physical, and dynamic parameters need to be considered in the scaling processes. We note that Assis et al. \cite{Assis, Assis2, Assis3} considered, besides the initial size ratio, parameters linked to the fluid flow and grain entrainment. We note also that differences in the microscopic (grain-scale) dynamics are expected. In the subaqueous case, grains move mainly by rolling and sliding, following closely the fluid flow and being susceptible to small vortices and other small flow structures. This has been shown to be especially important for the grains migrating toward the barchan horns \cite{Alvarez3, Alvarez4}. In the aeolian case, grains move by saltation and reptation, and those in saltation follow approximately straight trajectories in the main flow direction \cite{Bagnold_1}, being much less affected by small flow structures. Therefore, extrapolations of experimental results obtained under water to the aeolian case must be carried out with caution.

\section{Barchan-obstacle interactions}

Bacik et al. \cite{Bacik2} carried out the first systematic investigation of dune-obstacle interactions, for quasi-2D dunes. By using the annular flume described in Subsection \ref{section:methods}\ref{subsection:experiments},  in which different 2D obstacles were placed on the bottom, they found that 2D dunes either cross over the obstacle or remain trapped depending on the relative size of the obstacle with respect to the dune: large obstacles favor trapping, while small favor crossing. They showed that the dune behavior is controlled by the water flow structure near the obstacle (extended wake regions and downstream recirculation zones enhance sediment retention and, thus, promote trapping). Their results paved the way for later investigations on 3D dunes.

\begin{figure}[ht]
	\centering
	\includegraphics[width=\linewidth]{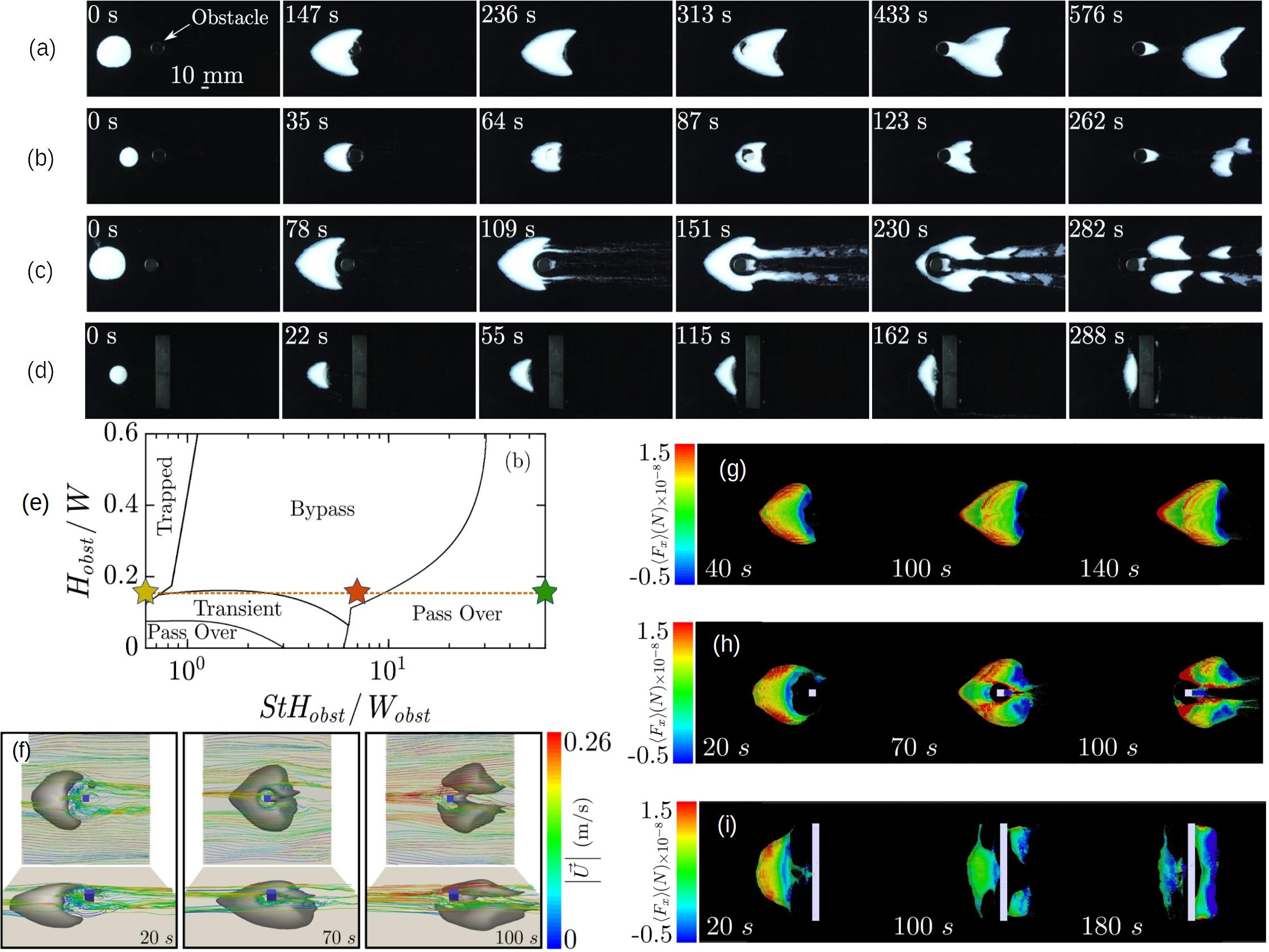}
	\caption{Snapshots from experiments of a barchan dune interacting with an obstacle, for different size ratios and shapes. In the snapshots, the water flow is from left to right, and the corresponding times are shown in each frame. (a) Cylinder with $H_{obst}/W$ $=$ 0.05, for which the dune passes over the obstacle and continues as a single barchan (pass over pattern); (b) cylinder with $H_{obst}/W$ $=$ 0.08, for which the dune behaves in an intermediate situation between passing over and bypassing the obstacle (transient case); (c) Cylinder with $H_{obst}/W$ $=$ 0.19, for which the dune bypasses the obstacle; (d) block with $H_{obst}/W$ = 0.40 and $H_{obst}/W_{obst}$ = 0.10, for which the dune is trapped and spreads in the spanwise direction. (e) Classification map in the size ratio $H_{obst}/W$ vs. modified Stokes number St$H_{obst} / W_{obst}$ apace, proposed by Assis et al. \cite{Assis4}, with the three points simulated by Lima et al. \cite {Lima3}. (f) Instantaneous streamlines from numerical simulations at different stages of the barchan-obstacle interaction, for the bypass case. (g-i) Snapshots showing top view images of a dune colored in accordance with the average level of the longitudinal component of resultant force of individual grains $\left< F_x \right>$: (g) pass over; (h) bypass; and (i) trapped cases. Panels (a-d) were extracted from Assis et al. \cite{Assis4}, https://doi.org/10.1029/2023GL104125, and panels (e-i) from Lima et al. \cite {Lima3}, https://doi.org/10.1029/2025JF008504.}
	\label{fig:dune_obstacle}
\end{figure}

Assis et al. \cite{Assis4} examined the interaction between subaqueous barchans and dune-scale obstacles through water-channel experiments (setup described in Subsection \ref{section:methods}\ref{subsection:experiments}) in which a large three-dimensional obstacle was initially placed downstream of an isolated barchan. By varying the obstacle geometry and size, flow velocity, and grain properties, they showed that barchans may be blocked (trapped, Fig. \ref{fig:dune_obstacle}d), circumvent (bypass, Fig. \ref{fig:dune_obstacle}c) the obstacle, or pass over it (Fig. \ref{fig:dune_obstacle}a), with transitional behaviors (Fig. \ref{fig:dune_obstacle}b) occurring between these regimes. Interestingly, when subaqueous barchans bypass the obstacle, their grains do not touch the latter and a region with virtually no grains exists in the vicinity of the obstacle (as seen in Fig. \ref{fig:dune_obstacle}c). Based on these observations, Assis et al. (2023) proposed an \textit{ad hoc} classification map in which the interaction outcome is governed by two dimensionless parameters: a modified Stokes number,

\begin{equation}
	\mathrm{St} \frac{H_{obst}}{W_{obst}} \,\,,
\end{equation}

\noindent and the obstacle-to-dune size ratio,

\begin{equation}
	\frac{H_{obst}}{W} \,\,.
\end{equation}

\noindent The results are well organized in the $H_{obst}/W$ vs. St$ H_{obst} / W_{obst}$ space, shown in Fig. \ref{fig:dune_obstacle}e. In principle, these maps provide a predictive framework for determining the outcome of barchan–obstacle interactions.

More recently, Lima et al. \cite{Lima3} extended the investigation of Assis et al. \cite{Assis4} by using CFD–DEM simulations, providing a mechanistic explanation for the observed interaction patterns. Their simulations resolved both the fluid and granular flows at the grain scale, allowing direct measurement of grain trajectories, resultant forces on each particle, and local erosion and deposition rates during interaction with an obstacle. Among other results, they showed that in the pass over regime flow perturbations induced by the obstacle are weak, so that grains in the central region of the dune follow nearly straight trajectories that move over the obstacle, while grains near the dune periphery move along curved paths, as also observed for isolated barchans \cite{Alvarez3}. In the trapped regime, grains follow closely the fluid pathlines, which are approximately circular, which means that some particles remain confined within the recirculation zone downstream of the obstacle, whereas others are entrained further downstream by the fluid flow. In the bypass and trapped cases, the results show the existence of a strong vortex between the dune and the obstacle, being the result of interactions between the recirculation region downstream the dune and a horseshoe vortex that exists upstream the obstacle. The resulting vortex has enough strength to deviate the main flow and hinder grains from touching or passing over the obstacle. Locally, the obstacle redistributes the basal shear stress and modifies recirculation patterns, thereby altering the sediment flux. Finally, Lima et al. \cite{Lima3} show the distribution of the resultant force acting on grains, which is consistent with the grain trajectories found in simulations and experiments.

Taken together, these studies establish a coherent picture in which dune-obstacle interactions are governed by a balance between flow perturbation, grain-scale transport, and geometric constraints.

\subsection{Implications of subaqueous findings for aeolian barchans}

In subaqueous environments, the density ratio between grains and fluid is relatively low ($S$ $=$ 2.5, compared to $S$ $\approx$ 2500 in aeolian conditions), so that sediment transport occurs primarily as bedload, with grains rolling and sliding while remaining in contact with a static or creeping layer. As a consequence, grains follow closely the fluid flow and are susceptible to small vortices and other local flow structures, which are known to play an important role in the migration of grains toward the barchan horns \cite{Alvarez3, Alvarez4}. When a strong vortex forms around an obstacle and the flow accelerates along its flanks, the grains are diverted by the fluid and tend to follow trajectories closely aligned with the local streamlines, thereby avoiding direct contact with the obstacle. In addition, in the trapped and bypass regimes observed in subaqueous experiments, vortical structures generated by the interaction between the dune wake and the horseshoe vortex upstream of the obstacle create regions depleted of grains, strongly affecting the morphodynamics of the interaction.

The situation contrasts with the aeolian case, where grains possess significant inertia and move predominantly by saltation and reptation. In particular, grains in saltation describe ballistic trajectories that are much less constrained by instantaneous flow deviations and small-scale flow structures \cite{Bagnold_1}. Under such conditions, particles can readily impact obstacles even when the airflow is deflected around them, which explains the frequent accumulation of sand upstream of buildings and other structures in desert regions. For relatively high obstacles, however, scour regions may develop upstream because of the presence of horseshoe vortices, as observed, for example, around the pyramids of Mero\"e in Sudan \cite{Raffaele}. Therefore, in the context of barchan-obstacle interactions, the pronounced void regions observed in subaqueous bypass and trapping regimes are unlikely to develop with the same intensity in aeolian settings.

More generally, although subaqueous experiments reproduce many morphodynamic features of aeolian barchans and constitute an important laboratory model for investigating dune interactions, important differences remain at the grain scale because of the different transport modes and grain-fluid coupling. Thus, although subaqueous results contribute to the understanding of dune-obstacle interactions, their direct extrapolation to aeolian environments should be made with caution. Beyond its fundamental relevance to morphodynamic stability, this problem bears direct implications for predicting scour, burial, and structural exposure in hydraulic and offshore infrastructures, providing a physically grounded framework for the design of resilient and environmentally sustainable systems.

\section{Conclusions and perspectives}

In this paper, we reviewed the state-of-the-art of barchan–barchan and barchan–obstacle interactions, discussing the main results acquired from remote sensing, laboratory experiments, and numerical simulations. From different methodologies and environments, a consistent picture emerges: barchans act as interacting systems in which mass redistribution, flow perturbations, and geometric constraints collectively regulate dune size, spacing, and migration. Rather than evolving as isolated bedforms, barchans continuously exchange sediment and momentum through collisions and wake-induced regimes, and interactions with fixed obstacles. These processes contribute to the self-organization of dune corridors and persistence of relatively narrow barchan-size distributions observed in dune fields.

Despite significant progress, several questions remain open. First, most laboratory and numerical studies focus on binary interactions under steady forcing, whereas natural dune fields experience fluctuating winds and multi-body interactions; extending controlled studies to include unsteadiness and collective effects is essential. Second, while continuum and agent-based models successfully capture large-scale organization, and CFD–DEM resolves grain-scale dynamics, a unified framework linking microscale transport to field-scale pattern formation is still lacking. Third, dune-obstacle interactions in aeolian environments are comparatively less explored, and systematic field-based validation of predictive maps derived from subaqueous experiments is still necessary. Therefore, some fronts to be explored in the near future are (i) experiments and numerical simulations on multi-dunes; (ii) barchan-barchan interactions under oscillating flows; (iii) mechanistic or stochastic models to solve statistics at the grain scale that affect the dynamics at the dune or dune-field scales; and (iv) mapping barchan-obstacle and more barchan-barchan interactions in aeolian environment. Advancing these fronts will improve our ability to predict the evolution of dune fields, coping with the expansion of deserts and managing hazards in regions where migrating dunes interact with infrastructure.

\vskip6pt

\ack{E.M.F. is grateful to FAPESP (Grant Nos. 2018/14981-7 and 2025/20488-5) and to CNPq (Grant No. 405512/2022-8) for the financial support provided. The authors are grateful to Esteban C\'u\~nez for discussions on the use of the Copernicus platform.}

\bibliographystyle{plain} 
\bibliography{references}

\end{document}